\makeatletter

\font\tenmsx=msxm10
\font\sevenmsx=msxm7
\font\fivemsx=msxm5
\font\tenmsy=msym10
\font\sevenmsy=msym7
\font\fivemsy=msym5
\newfam\msxfam
\newfam\msyfam
\textfont\msxfam=\tenmsx  \scriptfont\msxfam=\sevenmsx
  \scriptscriptfont\msxfam=\fivemsx
\textfont\msyfam=\tenmsy  \scriptfont\msyfam=\sevenmsy
  \scriptscriptfont\msyfam=\fivemsy

\def\hexnumber@#1{\ifnum#1<10 \number#1\else
 \ifnum#1=10 A\else\ifnum#1=11 B\else\ifnum#1=12 C\else
 \ifnum#1=13 D\else\ifnum#1=14 E\else\ifnum#1=15 F\fi\fi\fi\fi\fi\fi\fi}

\def\msx@{\hexnumber@\msxfam}
\def\msy@{\hexnumber@\msyfam}
\mathchardef\boxdot="2\msx@00
\mathchardef\boxplus="2\msx@01
\mathchardef\boxtimes="2\msx@02
\mathchardef\square="0\msx@03
\mathchardef\blacksquare="0\msx@04
\mathchardef\centerdot="2\msx@05
\mathchardef\lozenge="0\msx@06
\mathchardef\blacklozenge="0\msx@07
\mathchardef\circlearrowright="3\msx@08
\mathchardef\circlearrowleft="3\msx@09
\mathchardef\rightleftharpoons="3\msx@0A
\mathchardef\leftrightharpoons="3\msx@0B
\mathchardef\boxminus="2\msx@0C
\mathchardef\Vdash="3\msx@0D
\mathchardef\Vvdash="3\msx@0E
\mathchardef\vDash="3\msx@0F
\mathchardef\twoheadrightarrow="3\msx@10
\mathchardef\twoheadleftarrow="3\msx@11
\mathchardef\leftleftarrows="3\msx@12
\mathchardef\rightrightarrows="3\msx@13
\mathchardef\upuparrows="3\msx@14
\mathchardef\downdownarrows="3\msx@15
\mathchardef\upharpoonright="3\msx@16

\mathchardef\downharpoonright="3\msx@17
\mathchardef\upharpoonleft="3\msx@18
\mathchardef\downharpoonleft="3\msx@19
\mathchardef\rightarrowtail="3\msx@1A
\mathchardef\leftarrowtail="3\msx@1B
\mathchardef\leftrightarrows="3\msx@1C
\mathchardef\rightleftarrows="3\msx@1D
\mathchardef\Lsh="3\msx@1E
\mathchardef\Rsh="3\msx@1F
\mathchardef\rightsquigarrow="3\msx@20
\mathchardef\leftrightsquigarrow="3\msx@21
\mathchardef\looparrowleft="3\msx@22
\mathchardef\looparrowright="3\msx@23
\mathchardef\circeq="3\msx@24
\mathchardef\succsim="3\msx@25
\mathchardef\gtrsim="3\msx@26
\mathchardef\gtrapprox="3\msx@27
\mathchardef\multimap="3\msx@28
\mathchardef\therefore="3\msx@29
\mathchardef\because="3\msx@2A
\mathchardef\doteqdot="3\msx@2B

\mathchardef\triangleq="3\msx@2C
\mathchardef\precsim="3\msx@2D
\mathchardef\lesssim="3\msx@2E
\mathchardef\lessapprox="3\msx@2F
\mathchardef\eqslantless="3\msx@30
\mathchardef\eqslantgtr="3\msx@31
\mathchardef\curlyeqprec="3\msx@32
\mathchardef\curlyeqsucc="3\msx@33
\mathchardef\preccurlyeq="3\msx@34
\mathchardef\leqq="3\msx@35
\mathchardef\leqslant="3\msx@36
\mathchardef\lessgtr="3\msx@37
\mathchardef\backprime="0\msx@38
\mathchardef\risingdotseq="3\msx@3A
\mathchardef\fallingdotseq="3\msx@3B
\mathchardef\succcurlyeq="3\msx@3C
\mathchardef\geqq="3\msx@3D
\mathchardef\geqslant="3\msx@3E
\mathchardef\gtrless="3\msx@3F
\mathchardef\sqsubset="3\msx@40
\mathchardef\sqsupset="3\msx@41
\mathchardef\trianglerighteq="3\msx@44
\mathchardef\trianglelefteq="3\msx@45
\mathchardef\bigstar="0\msx@46
\mathchardef\between="3\msx@47
\mathchardef\blacktriangledown="0\msx@48
\mathchardef\blacktriangleright="3\msx@49
\mathchardef\blacktriangleleft="3\msx@4A
\mathchardef\blacktriangle="0\msx@4E
\mathchardef\triangledown="0\msx@4F
\mathchardef\eqcirc="3\msx@50
\mathchardef\lesseqgtr="3\msx@51
\mathchardef\gtreqless="3\msx@52
\mathchardef\lesseqqgtr="3\msx@53
\mathchardef\gtreqqless="3\msx@54
\mathchardef\Rrightarrow="3\msx@56
\mathchardef\Lleftarrow="3\msx@57
\mathchardef\veebar="2\msx@59
\mathchardef\barwedge="2\msx@5A
\mathchardef\doublebarwedge="2\msx@5B
\mathchardef\angle="0\msx@5C
\mathchardef\measuredangle="0\msx@5D
\mathchardef\sphericalangle="0\msx@5E
\mathchardef\varpropto="3\msx@5F
\mathchardef\smallsmile="3\msx@60
\mathchardef\smallfrown="3\msx@61
\mathchardef\Subset="3\msx@62
\mathchardef\Supset="3\msx@63
\mathchardef\Cup="2\msx@64

\mathchardef\Cap="2\msx@65

\mathchardef\curlywedge="2\msx@66
\mathchardef\curlyvee="2\msx@67
\mathchardef\leftthreetimes="2\msx@68
\mathchardef\rightthreetimes="2\msx@69
\mathchardef\subseteqq="3\msx@6A
\mathchardef\supseteqq="3\msx@6B
\mathchardef\bumpeq="3\msx@6C
\mathchardef\Bumpeq="3\msx@6D
\mathchardef\lll="3\msx@6E

\mathchardef\ggg="3\msx@6F

\mathchardef\circledS="0\msx@73
\mathchardef\pitchfork="3\msx@74
\mathchardef\dotplus="2\msx@75
\mathchardef\backsim="3\msx@76
\mathchardef\backsimeq="3\msx@77
\mathchardef\complement="0\msx@7B
\mathchardef\intercal="2\msx@7C
\mathchardef\circledcirc="2\msx@7D
\mathchardef\circledast="2\msx@7E
\mathchardef\circleddash="2\msx@7F
\def\ulcorner{\delimiter"4\msx@70\msx@70 }
\def\urcorner{\delimiter"5\msx@71\msx@71 }
\def\llcorner{\delimiter"4\msx@78\msx@78 }
\def\lrcorner{\delimiter"5\msx@79\msx@79 }
\def\yen{\mathhexbox\msx@55 }
\def\checkmark{\mathhexbox\msx@58 }
\def\circledR{\mathhexbox\msx@72 }
\def\maltese{\mathhexbox\msx@7A }
\mathchardef\lvertneqq="3\msy@00
\mathchardef\gvertneqq="3\msy@01
\mathchardef\nleq="3\msy@02
\mathchardef\ngeq="3\msy@03
\mathchardef\nless="3\msy@04
\mathchardef\ngtr="3\msy@05
\mathchardef\nprec="3\msy@06
\mathchardef\nsucc="3\msy@07
\mathchardef\lneqq="3\msy@08
\mathchardef\gneqq="3\msy@09
\mathchardef\nleqslant="3\msy@0A
\mathchardef\ngeqslant="3\msy@0B
\mathchardef\lneq="3\msy@0C
\mathchardef\gneq="3\msy@0D
\mathchardef\npreceq="3\msy@0E
\mathchardef\nsucceq="3\msy@0F
\mathchardef\precnsim="3\msy@10
\mathchardef\succnsim="3\msy@11
\mathchardef\lnsim="3\msy@12
\mathchardef\gnsim="3\msy@13
\mathchardef\nleqq="3\msy@14
\mathchardef\ngeqq="3\msy@15
\mathchardef\precneqq="3\msy@16
\mathchardef\succneqq="3\msy@17
\mathchardef\precnapprox="3\msy@18
\mathchardef\succnapprox="3\msy@19
\mathchardef\lnapprox="3\msy@1A
\mathchardef\gnapprox="3\msy@1B
\mathchardef\nsim="3\msy@1C
\mathchardef\napprox="3\msy@1D
\mathchardef\nsubseteqq="3\msy@22
\mathchardef\nsupseteqq="3\msy@23
\mathchardef\subsetneqq="3\msy@24
\mathchardef\supsetneqq="3\msy@25
\mathchardef\subsetneq="3\msy@28
\mathchardef\supsetneq="3\msy@29
\mathchardef\nsubseteq="3\msy@2A
\mathchardef\nsupseteq="3\msy@2B
\mathchardef\nparallel="3\msy@2C
\mathchardef\nmid="3\msy@2D
\mathchardef\nshortmid="3\msy@2E
\mathchardef\nshortparallel="3\msy@2F
\mathchardef\nvdash="3\msy@30
\mathchardef\nVdash="3\msy@31
\mathchardef\nvDash="3\msy@32
\mathchardef\nVDash="3\msy@33
\mathchardef\ntrianglerighteq="3\msy@34
\mathchardef\ntrianglelefteq="3\msy@35
\mathchardef\ntriangleleft="3\msy@36
\mathchardef\ntriangleright="3\msy@37
\mathchardef\nleftarrow="3\msy@38
\mathchardef\nrightarrow="3\msy@39
\mathchardef\nLeftarrow="3\msy@3A
\mathchardef\nRightarrow="3\msy@3B
\mathchardef\nLeftrightarrow="3\msy@3C
\mathchardef\nleftrightarrow="3\msy@3D
\mathchardef\divideontimes="2\msy@3E
\mathchardef\varnothing="0\msy@3F
\mathchardef\nexists="0\msy@40
\mathchardef\mho="0\msy@66
\mathchardef\thorn="0\msy@67
\mathchardef\beth="0\msy@69
\mathchardef\gimel="0\msy@6A
\mathchardef\daleth="0\msy@6B
\mathchardef\lessdot="3\msy@6C
\mathchardef\gtrdot="3\msy@6D
\mathchardef\ltimes="2\msy@6E
\mathchardef\rtimes="2\msy@6F
\mathchardef\shortmid="3\msy@70
\mathchardef\shortparallel="3\msy@71
\mathchardef\smallsetminus="2\msy@72
\mathchardef\thicksim="3\msy@73
\mathchardef\thickapprox="3\msy@74
\mathchardef\approxeq="3\msy@75
\mathchardef\succapprox="3\msy@76
\mathchardef\precapprox="3\msy@77
\mathchardef\curvearrowleft="3\msy@78
\mathchardef\curvearrowright="3\msy@79
\mathchardef\digamma="0\msy@7A
\mathchardef\varkappa="0\msy@7B
\mathchardef\hslash="0\msy@7D
\mathchardef\hbar="0\msy@7E
\mathchardef\backepsilon="3\msy@7F
\def\Bbb{\ifmmode\let\next\Bbb@\else
 \def\next{\errmessage{Use \string\Bbb\space only in math mode}}\fi\next}
\def\Bbb@#1{{\Bbb@@{#1}}}
\def\Bbb@@#1{\fam\msyfam#1}

\def\inv{^{\raise.15ex\hbox{${
  \scriptscriptstyle -}$}\kern-.05em 1}}

\def\Dsl{\,\raise.15ex\hbox{$/$}\mkern-13.5mu D}
\def\dsl{\raise.15ex\hbox{$/$}\kern-.57em\hbox{$\partial$}}

\def\lspace{\ifx\answ\bigans{}\else\qquad\fi}
\def\del{\partial}

\def\lform{\hbox{$\sqcup$}\llap{\hbox{$\sqcap$}}}
\def\darr#1{\raise1.5ex\hbox{$\leftrightarrow$}
\mkern-16.5mu #1}

\def\INT{{\textstyle \int\kern-.642em\int}}

\def\R{{\Bbb R}}
\def\C{{\Bbb C}}
\def\Z{{\Bbb Z}}

\def\eps{{\epsilon}}

\def\cocross{{>\!\!\!\triangleleft}}

\def\tens{\mathop{\otimes}}

\def\la{{\triangleright}}

\def\id{{\rm id}}

\def\nquad{{\!\!\!\!\!\!}}
\def\nqquad{\nquad\nquad}
\def\eqn#1#2{\begin{equation}#2\label{#1}\end{equation}}

\def\haj#1{{\mathaccent20 {#1}}}
\def\Vhaj{{V\haj{\ }}}

\def\und#1{{\underline {#1}}}

\def\text#1{\mbox{\rm #1}}
\def\note#1{}

\def\blacksquare{{\lform}}
\def\frac#1#2{{{#1\over#2}}}

\def\proof{\goodbreak\noindent{\bf Proof\quad}}

\def\endproof{{\ $\lform$}\bigskip }

\def\align#1{\begin{eqnarray*}#1\end{eqnarray*}}

\def\alignn#1#2{\begin{eqnarray}\label{#1}#2
\end{eqnarray}}

\def\und#1{{\underline{#1}}}

\def\vecv{{\bf v}}
\def\vecx{{\bf x}}\def\vecp{{\bf p}}
\def\veca{{\bf a}}

\def\<{\langle}
\def\>{\rangle}

\makeatother

\documentstyle[11pt,epsf]{article}
\newtheorem{lemma}{Lemma}[section]
\newtheorem{propos}[lemma]{Proposition}
\newtheorem{example}[lemma]{Example}
\newtheorem{theorem}[lemma]{Theorem}

\newtheorem{corol}[lemma]{Corollary}
\newtheorem{defin}[lemma]{Definition}

\begin{document}\baselineskip 25pt

{\ }\hskip 4.7in DAMTP/93-3 
\vspace{.5in}

\begin{center}  {\Large FREE BRAIDED DIFFERENTIAL CALCULUS,}\\ {\Large BRAIDED
BINOMIAL THEOREM}\\ {\Large AND THE BRAIDED EXPONENTIAL MAP}
\\ \baselineskip 13pt{\ }
{\ }\\ S. Majid\footnote{SERC Fellow and Fellow of Pembroke College,
Cambridge}\\ {\ }\\
Department of Applied Mathematics\\
\& Theoretical Physics\\ University of Cambridge\\ Cambridge CB3 9EW, U.K.
\end{center}

\begin{center}
January 1993\end{center}
\vspace{10pt}
\begin{quote}\baselineskip 13pt

\noindent{\bf ABSTRACT}  Braided differential operators $\del^i$ are obtained
by differentiating the
addition law on the braided covector spaces introduced previously (such as the
braided addition law on the quantum plane). These are affiliated to a
Yang-Baxter matrix $R$. The quantum eigenfunctions $\exp_R(\vecx|\vecv)$ of the
$\del^i$ (braided-plane waves) are introduced in the free case where the
position components $x_i$ are totally non-commuting. We prove a braided
$R$-binomial theorem and a braided-Taylors theorem
$\exp_R(\veca|\del)f(\vecx)=f(\veca+\vecx)$. These various results precisely
generalise to a generic $R$-matrix (and hence to $n$-dimensions) the well-known
properties of the usual 1-dimensional $q$-differential and $q$-exponential. As
a related application, we show that the q-Heisenberg algebra $px-qxp=1$ is a
braided semidirect product $\C[x]\cocross \C[p]$ of the braided line acting on
itself (a braided Weyl algebra). Similarly for its generalization to an
arbitrary $R$-matrix.

\end{quote}
\baselineskip 22pt

\section{Introduction}

In recent times there has been a lot of interest in the programme of
$q$-deforming physics. This programme is
usually phrased in the language on non-commutative geometry, namely one works
with non-commutative algebras such as q-deformed planes, q-Minkowski space etc
and tries to proceed as if these algebras are like the commutative algebra of
functions on planes, Minkowski space etc. One would like to make all basic
constructions needed for physics in this q-deformed context. Here we develop
further our new and fully systematic approach to this problem. Instead of
working in a usual way but with non-commutative algebras, our new approach is
to introduce $q$ directly into the notion of tensor product. In the simplest
case $\tens$ acquires a factor $q$ in a similar way to the $\pm1$ factors
inserted in super-symmetry, while more generally the modification of the
$\tens$ is given by an $R$-matrix (which could depend on one or more
parameters). We will see that this approach recovers such things as the usual
$q$-differential and $q$-exponential, and the $q$-Heisenberg algebra in the
simplest case, but works just as well for any $R$-matrix.

Mathematically, our approach to $q$-deformations consists in shifting the
category in which we work from the category of vector spaces to a braided
tensor category. We have introduced groups and other structures in such braided
categories in
\cite{Ma:exa}\cite{Ma:bra}\cite{Ma:csta}\cite{Ma:lin}\cite{Ma:poi} and we build
on this work here. This line of development is a little different from
non-commutative geometry in that it is the tensor product that is being
primarily deformed (other deformations follow from this). Rather it extends the
philosophy of super-geometry to some kind of braided geometry.

A second physical motivation for braided geometry is the existence in low
dimensions of particles with braid statistics, whose symmetries might be
described in such a way. We will not develop this here, limiting ourselves to
the motivation from q-deforming physics. In q-deformed physics q is not a
physical parameter but a parameter introduced to help regularise
infinities\cite{Ma:reg}, and we can set $q=1$ after intelligent renormalization
(using identities from $q$-analysis). Perhaps in Planck-scale physics we might
also keep $q\ne 1$ as a model of quantum corrections to the geometry. In the
first case then the usual spin-statistics theorem which fails in the
$q$-deformed setting, does so as an artifact of the regularization process.

Since our results generalise these ideas to any generic $R$-matrix obeying the
Quantum Yang-Baxter Equations (QYBE) they are not of course tied to this
deformation point of view. There are plenty of non-standard solutions of the
QYBE to which our results apply equally well. We hope in this context to
provide some kind of differential calculus for the handling and computation of
the knot invariants and 3-manifold invariants associated to the chosen $R$
matrix (cf Fox's free calculus). This is a long-term pure-mathematical
motivation for the paper.

The paper begins in the preliminary Section~2 by recalling our approach to
braided-differential calculus announced in \cite[Sec. 7.3]{Ma:introp}. We begin
with an addition law on the quantum planes associated to an $R$-matrix, and
define differentiation as an infinitesimal translation. There is more than one
plane associated to an $R$-matrix. As well as ones associated to non-zero
`eigenvalues' of $R$ there is the free one in which the coordinates $x_i$ obey
no commutation relations at all. The first main result is in Section~3 where we
prove
a braided-binomial theorem precisely generalising the familiar (and
indispensable) q-binomial theorem. We then use this in Section~4 when we
introduce and study an $R$-exponential map in the free case. The problem of an
$R$-exponential map in the non-free case is left open.

In section~5 we turn to another application of the braided group technology.
Because Hopf algebras in braided categories (like familiar quantum groups) are
group-like objects one can make semidirect products\cite{Ma:bos}. In
particular, the differential operators $\del^i$ act on the $x_j$ and hence we
can make a semidirect product. In the usual setting this is precisely the
definition of the canonical commutation relations algebra (or Weyl algebra).
Thus we have at once from  the general theory in \cite{Ma:bos} a braided Weyl
algebra. In the one-dimensional case we recover the q-Heisenberg algebra
proposed for q-deformed quantum mechanics in \cite{Man:not}. See also
\cite{Ros:hei}\cite{SchWes:def}. In the higher-dimensional case we recover
something like the algebra studied in \cite{Kem:sym} and elsewhere for the
$SL_q(n)$ $R$-matrix (though with $x,p$ rather than $a,a^\dagger$). All of
these are thereby generalised to the case of a general quantum plane based on a
general $R$-matrix and its projectors. Most importantly, these algebras are
obtained now in a systematic way by a standard semidirect product construction,
requiring us only to remember the braid statistics.

We will not discuss it explicitly, but the category in which all our
constructions take place can be understood as the category of
(co)representations of a quantum group, at least for nice $R$. This means that
all our constructions are quantum-group covariant under this background quantum
group symmetry. Related to this, there is also a matrix braided group
$B(R)$\cite{Ma:exa} acting on our braided covectors and braided vectors. See
\cite{Ma:lin} for an elaboration of these points.

\section{Derivation of Braided-Differential Calculus}

We begin by recalling the notion of a braided group or Hopf algebra in a
braided category as introduced previously in
\cite{Ma:exa}\cite{Ma:bra}\cite{Ma:lin}\cite{Ma:poi} and elsewhere. This is
$(B,\und\Delta,\und S,\und\eps,\Psi)$ where $B$ is an algebra, $\Psi:B\tens
B\to B\tens B$ is a braiding (and obeys the QYBE) and $\und\Delta:B\to
B\und\tens B$ the braided-coproduct, $\und\eps:B\to k$ the braided counit and
$\und S:B\to B$. Everything we do in this paper works at an algebraic level
over any field $k$ (one can take $k=\R,\C$). For full details of the axioms see
the above works or \cite{Ma:introp}. They are just the same as the usual Hopf
algebra axioms except that $\und\Delta$ is an algebra homomorphism when
$B\und\tens B$ has the {\em braided tensor product} algebra structure
\eqn{bratens}{ (a\tens c)(b\tens d)=a\Psi(c\tens b)d}
as determined by $\Psi$. The braiding on the identity is always trivial. In
more formal terms, the construction takes place in a general braided tensor
category\cite{Ma:eul}\cite{Ma:bg}.

The example which will concern us here is the quantum plane and its
higher-dimensional brethren. These have been extensively studied as algebras
but without any linear addition law. This is because to formulate the latter
one
really needs the notion of a braided-Hopf algebra. In fact, there are two
flavours, contravariant and covariant (covector and vector) associated to any
general $R$ matrix. Let $R\in M_n\tens M_n$ obey the QYBE. We suppose it is
invertible. We suppose also that we are given an affiliated matrix $R'$ such
that
\eqn{R'}{ (PR+1)(PR'-1)=0,\quad R_{12}R_{13}R'_{23}=R'_{23}R_{13}R_{12},\
R_{23}R_{13}R'_{12}=R'_{12}R_{13}R_{23},\  R_{21}R'_{12}=R'_{21}R_{12}}
where $P$ is the usual permutation matrix. Here $PR'-1$ corresponds to the
projection operators in the usual approach to the algebra in
\cite{OSWZ:def}\cite{CWSSW:lor} and elsewhere. One choice (which we call the
free case) is $R'=P$. Other $R'$ are associated to each non-zero $\lambda_i$ in
the characteristic equation $\Pi_i(PR-\lambda_i)=0$ (normalise $R$ so that the
chosen eigenvalue becomes $-1$ and construct $R'$ from the remaining products
and $P$).

With this data the {\em braided covectors} $\Vhaj(R')$ defined by generators
$1,\vecx=(x_i)$ and relations $\vecx_2\vecx_1R'_{12}=\vecx_1\vecx_2$ form a
braided-Hopf algebra with
\eqn{covecdelta}{ \und\Delta x_i=x_i\tens 1+1\tens x_i,\qquad \Psi(\vecx_1\tens
\vecx_2)=\vecx_2\tens \vecx_1R_{12}}
extended multiplicatively with braid statistics. The counit is $\eps(x_i)=0$.
The antipode is $\und S x_i=-x_i$ extended as a braided-antialgebra map.
Likewise there are {\em braided vectors} $V(R')$ defined by
$R'_{12}\vecv_2\vecv_1=\vecv_1\vecv_2$ form a braided-Hopf algebra with
\eqn{vecdelta}{\und\Delta v^i=v^i\tens 1+1\tens v^i,\qquad \Psi(\vecv_1\tens
\vecv_2)=R_{12}\vecv_2\tens \vecv_1.}
We refer to \cite{Ma:poi} for full details. Throughout the paper we use the
standard compact notation in which numerical suffices on vectors, covectors and
matrices refer to the position in a matrix tensor product. In more
old-fashioned notation the covector and vector relations are
\eqn{comprelns}{ x_{i_1}x_{i_2}=x_{j_2}x_{j_1}
R^{j_1}{}_{i_1}{}^{j_2}{}_{i_2},\quad
v^{i_1}v^{i_2}=R^{i_1}{}_{j_1}{}^{i_2}{}_{j_2} v^{j_2}v^{j_1}.}

We consider $\Vhaj(R')$ as something like the functions on $\R^n$ with $x_i$
something like the co-ordinate function that picks out the $i$-th component.
The linear braided-coproduct then corresponds to addition on $\R^n$. In
practice of course we work with $x_i$ directly. Also, the braided-coproduct
$\Vhaj(R')\to \Vhaj(R')\und\tens \Vhaj(R')$ can be regarded just as well as a
braided-coaction of one copy on the other. If we write $\veca\equiv \vecx\tens
1$ and $\vecx\equiv 1\tens\vecx$ for the generators of the two copies of
$\Vhaj(R')$ then  $\Vhaj(R')\und\tens \Vhaj(R')$ is generated by the two copies
and cross relations
\eqn{crossax}{\vecx_1\veca_2=\veca_2\vecx_1 R_{12}=\veca_1\vecx_2 (PR)_{12}}
as determined by $\Psi$ from (\ref{bratens}). The braided coproduct or coaction
is
\eqn{braadd}{\vecx\mapsto \veca+\vecx}
and says that we can add braided covectors (the sum also realises $\Vhaj(R')$
provided we remember the braid statistics (\ref{crossax})).

\begin{defin} We define the braided-differential operators $\del^i:\Vhaj(R')\to
\Vhaj(R')$ as
\[ \del^i f(\vecx)=
\left(a_i^{-1}(f(\veca+\vecx)-f(\vecx))\right)_{\veca=0}\equiv{\rm coeff\ of\
}a^i{\rm \ in\ } f(\veca+\vecx).\]
\end{defin}

In other words, the $\del^i$ are the generators of infinitesimal translations.
Remembering the braid statistics we have on the monomials
\alignn{di}{\del^i(\vecx_1\cdots \vecx_m)&=& {\rm coeff}_{a^i}
\left((\veca_1+\vecx_1)(\veca_2+\vecx_2)
\cdots(\veca_m+\vecx_m)\right)\nonumber\\
&=&{\rm coeff}_{a^i}
\left(\veca_1\vecx_2\cdots\vecx_m+\vecx_1\veca_2\vecx_3\cdots\vecx_m
+\cdots+\vecx_1\cdots\vecx_{m-1}\veca_m\right)\nonumber\\
&=& {\rm coeff}_{a^i}\left
(\veca_1\vecx_2\cdots\vecx_m(1+(PR)_{12}+(PR)_{12}(PR)_{23}
+\cdots+(PR)_{12}\cdots (PR)_{m-1, m})\right)\nonumber\\
&=& {\bf e}^i{}_1\vecx_2\cdots\vecx_m \left[m;R\right]_{1\cdots m}}
where ${\bf e}^i$ is a basis covector $({\bf e}^i){}_j=\delta^i{}_j$ and
\eqn{branum}{\left[m;R\right]=1+(PR)_{12}+(PR)_{12}(PR)_{23}
+\cdots+(PR)_{12}\cdots (PR)_{m-1,m}}
is a certain matrix living in the $m$-fold matrix tensor product. We call such
matrices {\em braided integers} for reasons that will become apparent in the
next section. In explicit terms we have
\eqn{dicomp}{ \del^ix_{i_1}\cdots x_{i_m}=\delta^i{}_{j_1}x_{j_2}\cdots
x_{j_m}\left[m;R\right]^{j_1\cdots j_m}_{i_1\cdots i_m}.}

\begin{propos} The operators $\del^i$ obey the relations of the braided vectors
$V(R')$. Thus there is an action $V(R')\tens \Vhaj(R')\to \Vhaj(R')$ given by
$v^i\tens f(\vecx)\mapsto \del^i f(\vecx)$.
\end{propos}
\proof We show the identity
\eqn{R'mm}{\left[m-1;R\right]_{2\cdots m}\left[m;R\right]_{1\cdots
m}=(PR')_{12}\left[m-1;R\right]_{2\cdots m}\left[m;R\right]_{1\cdots m}.}
To do this we use the definition of the braided-integers in (\ref{branum}), and
(\ref{R'}) to compute
\align{&&\nqquad ((PR')_{12}-1)\left[m-1;R\right]_{2\cdots
m}\left[m;R\right]_{1\cdots m}\\
&&= ((PR')_{12}-1)\left[m-1;R\right]_{2\cdots
m}\left(1+(PR)_{12}\left[m-1;R\right]_{2\cdots m}\right)\\
&&= ((PR')_{12}-1)\left((1+(PR)_{12})\left[m-1;R\right]_{2\cdots
m}+(\left[m-1;R\right]_{2\cdots m}-1)(PR)_{12} \left[m-1;R\right]_{2\cdots
m}\right)\\
&&= ((PR')_{12}-1)(PR)_{23}\left[m-2;R\right]_{3\cdots
m}(PR)_{12}\left[m-1;R\right]_{2\cdots m}\\
&&= ((PR')_{12}-1)(PR)_{23}(PR)_{12}\left[m-2;R\right]_{3\cdots
m}\left[m-1;R\right]_{2\cdots m}\\
&&=(PR)_{23}(PR)_{12}((PR')_{23}-1)\left[m-2;R\right]_{3\cdots
m}\left[m-1;R\right]_{2\cdots m}.}
Hence if we assume (\ref{R'mm}) for $m-1$ as an induction hypothesis, the last
expression vanishes. Hence the result also holds for $m$. The start of the
induction is provided by (\ref{R'}) itself.

{}From (\ref{di}) we have
\align{\del^i\del^k\vecx_1\cdots\vecx_m&=&{\bf e}^k{}_1{\bf
e}^i_2\vecx_3\cdots\vecx_m\left[m-1;R\right]_{2\cdots
m}\left[m;R\right]_{1\cdots m}\\
R'{}^i{}_a{}^k{}_b\del^b\del^a\vecx_1\cdots\vecx_m&
=&R'{}^i{}_a{}^k{}_b\del^b{\bf e}^a_1\vecx_2\cdots\vecx_m
\left[m;R\right]_{1\cdots m}\\
&=&R'{}^i{}_a{}^k{}_b{\bf e}^a_1{\bf
e}^b_2\vecx_3\cdots\vecx_m\left[m-1;R\right]_{2\cdots
m}\left[m;R\right]_{1\cdots m}\\
&=&{\bf e}^k{}_1{\bf e}^i{}_2\vecx_3\cdots\vecx_m
(PR')_{12}\left[m-1;R\right]_{2\cdots m}\left[m;R\right]_{1\cdots m}.}
These are equal due to our identity (\ref{R'mm}). Hence $\del^i$ obey the
relations for $v^i$ in (\ref{comprelns}) and are therefore an operator
realization of $V(R')$. \endproof

\begin{lemma} The operators $\del^i$ obey the braided-Leibniz rule
\[ \del^i(ab)=(\del^i a)b+\cdot\Psi^{-1}(\del^i\tens a)b\]
\end{lemma}
\proof We need here the inverse braiding between braided-vectors and
braided-covectors. This is computed from \cite{Ma:lin} and comes out as
$\Psi^{-1}(\del_2\tens \vecx_1)= \vecx_1\tens R_{12}\del_2 $ and extends to
products by functoriality as
\eqn{barpsidx}{\Psi^{-1}(\del^i\tens\vecx_1\cdots\vecx_r)={\bf
e}^i{}_1\vecx_2\cdots\vecx_r\vecx_{r+1}(PR)_{12}\cdots (PR)_{r,r+1}\del_{r+1}.}
Taking this in the proposition, we compute the right hand side on monomials as
\align{&&\nqquad
(\del^i\vecx_1\cdots\vecx_r)\vecx_{r+1}\cdots\vecx_{m}+\Psi^{-1}
(\del^i\tens\vecx_1\cdots\vecx_r)\vecx_{r+1}\cdots\vecx_m\\
&&={\bf e}^i{}_1\vecx_2\cdots\vecx_r\left[r;R\right]_{1\cdots
r}\vecx_{r+1}\cdots\vecx_m
+{\bf e}^i{}_1\vecx_2\cdots\vecx_r\vecx_{r'+1}(PR)_{12}\cdots
(PR)_{r,r'+1}\del_{r'+1}\vecx_{r+1}\cdots\vecx_m\\
&&={\bf
e}^i{}_1\vecx_2\cdots\vecx_r\vecx_{r+1}\cdots\vecx_m\left[r;R\right]_{1\cdots
r}
\\
&&\qquad\qquad +{\bf e}^i{}_1\vecx_2\cdots\vecx_r\vecx_{r+1}(PR)_{12}\cdots
(PR)_{r,r+1}\vecx_{r+2}\cdots\vecx_m  \left[m-r;R\right]_{r+1\cdots m}}
where we use (\ref{di}) to evaluate the differentials. The primed $r'+1$ labels
a distinct matrix space from the existing $r+1$ index. These are then
identified by the ${\bf e}_{r+1}$ brought down by the action of $\del_{r'+1}$.
The resulting expression coincides with ${\bf
e}^i{}_1\vecx_2\cdots\vecx_r\vecx_{r+1}\cdots \vecx_m \left[m;R\right]_{1\cdots
m}$
from the left hand side of the proposition, since
\eqn{addbranum}{\left[r;R\right]_{1\cdots r}
+(PR)_{12}\cdots (PR)_{r,r+1}\left[m-r;R\right]_{r+1\cdots m}=
\left[m;R\right]_{1\cdots m}.}
This is evident from the definition in (\ref{branum}). \endproof

\begin{example} Let
\[ R=\pmatrix{q^2&0&0&0\cr0&q&q^2-1&0 \cr 0&0&q&0\cr0&0&0&q^2},\qquad
R'=q^{-2}R\]
(the $SL_q(2)$ $R$-matrix). Then $\Vhaj(R')$ is the standard quantum plane
$\vecx=(x\ y)$ with relations $yx=qxy$ and $\del^i$ are
\[ {\del\over\del x} x^ny^m=\left[n;q^2\right] x^{n-1} y^m,\quad
{\del\over\del y} x^ny^m=q^{n}x^n\left[m;q^2\right] y^{m-1},\quad
\left[m;q^2\right]={q^{2m}-1\over q^2-1}.\]
\end{example}

This is similar to the results of another approach based on differential forms
in \cite{OSWZ:def} and elsewhere. Also, for any $R$-matrix we can take $R'=P$.
Then $\Vhaj(P)=k<x_i>$ is the free non-commutative algebra in $n$
indeterminates. $V(P)$ is likewise free and $\del^i$ are given from (\ref{di}).
We call this the {\em free braided differential calculus} associated to an
$R$-matrix. It is in a certain sense universal.

\begin{example}\cite{Ma:poi} Let $R=(q)$ the one-dimensional solution of the
QYBE. The free braided-line introduced in \cite{Ma:csta} is $\Vhaj(R')=k[x]$
(polynomials in one variable) with braiding $\Psi(x\tens x)=q x\tens x$. In
this case we have
\[\del x^m=x^{m-1}(1+q+\cdots+q^{m-1})=\left[m;q\right]
x^{m-1}\quad\Rightarrow\quad \del f(x)={f(qx)-f(x)\over (q-1)x}.\]
\end{example}

One can verify the various properties in Proposition~2.2 and Lemma~2.3  easily
enough in these examples. In the braided-line case the braided-Leibniz rule is
easily verified in the form
\[ \del(x^{n}x^{m})=(\del x^n)x^m+q^{n}x^n(\del x^m)\]
which is just as in the case of a super-derivation, but with $q$ in the role of
$(-1)$ and a $\Z$-grading in the role of $\Z_2$-grading. The degree of $\del$
here is $-1$ and the degree of $x^n$ is $n$. Similarly in Example~2.4 the $q^n$
factor can be thought of as arising from the braided Leibniz rule as
$\del\over\del y$ passes $x^n$. The general case based on an $R$-matrix should
be viewed in just the same way. Of course products of the $\del^i$ do not obey
the same Leibniz rule, but rather one deduced from the one shown for the
generators. We also note that for a fully categorical picture we need the all
possible braidings $\Psi,\Psi^{-1}$ between vectors and covectors, not only the
limited combinations used above. For this one needs that $R^{t_2}$ is
invertible as well as $R$, where $t_2$ denotes transposition in the second
matrix factor. We come to this full picture in  Section~5.

\section{Braided Binomial Theorem}

In this section we prove an $R$-matrix version of the familiar $q$-binomial
theorem. This tells us how to expand $(a+x)^m$ when $xa=qax$ say. The
$q$-binomial theorem has $q$-integers $\left[r;q\right]$ in place of usual
integers. The $R$-matrix generalization covers both the non-commutativity which
takes the form (\ref{crossax}) and the fact that we deal with the
multidimensional case. We need this theorem in later section. We let $R\in
M_n\tens M_n$ be an invertible solution of the QYBE and $\left[m;R\right]$ the
braided-integers introduced in (\ref{branum}).

\begin{defin} Let $\left[{m\atop r};R\right]$ be the matrix defined recursively
by
\[ \left[{m\atop r};R\right]_{1\cdots m}=(PR)_{r, r+1}\cdots (PR)_{m-1,
m}\left[{m-1\atop r-1};R\right]_{1\cdots m-1}+\left[{m-1\atop
r};R\right]_{1\cdots m-1}\]
\[  \left[{m\atop 0};R\right]=1, \qquad \left[{m\atop r};R\right]=0\quad {\rm
if}\ r>m.\]
The suffices here refer to the matrix position in tensor powers of $M_n$ in the
standard way.
\end{defin}
This defines in particular
\eqn{mm}{\left[{m\atop m};R\right]=\left[{m-1\atop
m-1};R\right]=\cdots=\left[{1\atop 1};R\right]=1}
\eqn{m1}{ \left[{m\atop 1};R\right]=(PR)_{12}\cdots (PR)_{m-1,
m}+\left[{m-1\atop 1};R\right]=\cdots=\left[m;R\right].}
A similar recursion defines $\left[{m\atop 2};R\right]_{1\cdots m}$ in terms of
$ \left[{m\atop 1};R\right]$ (which is known) and $\left[{m-1\atop
2};R\right]$, and similarly (in succession) up to $r=m$.

\begin{propos} Suppose that $\vecx_1\veca_2=\veca_1\vecx_2 (PR)_{12}$ (as in
the braided tensor product algebra in Section~2). Then
\[ (\veca_1+\vecx_1)\cdots
(\veca_m+\vecx_m)=\sum_{r=0}^{r=m}\veca_1\cdots\veca_r\vecx_{r+1}
\cdots\vecx_m\left[{m\atop r};R\right]_{1\cdots m}.\]
\end{propos}
\proof We proceed by induction. Suppose the proposition is true for $m-1$, then
\align{(&&\nqquad \veca_1+\vecx_1)\cdots
(\veca_m+\vecx_m)=(\veca_1+\vecx_1)\cdots
(\veca_{m-1}+\vecx_{m-1})\veca_m+(\veca_1+\vecx_1)\cdots
(\veca_{m-1}+\vecx_{m-1})\vecx_m\\
&&=\sum_{r=0}^{m-1}\veca_1\cdots\veca_r\vecx_{r+1}\cdots
\vecx_{m-1}\veca_m\left[{m-1\atop r};R\right]_{1\cdots m-1}
+(\veca_1+\vecx_1)\cdots (\veca_{m-1}+\vecx_{m-1})\vecx_m\\
&&=\sum_{r=1}^{m}\veca_1\cdots\veca_{r-1}\vecx_{r}\cdots\vecx_{m-1}
\veca_m\left[{m-1\atop r-1};R\right]_{1\cdots m-1}
+\sum_{r=0}^{m-1}\veca_1\cdots\veca_r\vecx_{r+1}\cdots\vecx_m\left[{m-1\atop
r};R\right]_{1\cdots m-1}\\
&&=\sum_{r=1}^{m}\veca_1\cdots\veca_{r-1}\veca_r\vecx_{r+1}
\cdots\vecx_{m}(PR)_{r,r+1}\cdots (PR)_{m-1,m}
\left[{m-1\atop r-1};R\right]_{1\cdots m-1}\\
&&\qquad\qquad+\sum_{r=0}^{m-1}\veca_1\cdots\veca_r\vecx_{r+1}
\cdots\vecx_m\left[{m-1\atop r};R\right]_{1\cdots m-1}\\
&&=\sum_{r=1}^{m}\veca_1\cdots\veca_r\vecx_{r+1}\cdots\vecx_{m}\left[{m\atop
r};R\right]_{1\cdots m}
+\vecx_{1}\cdots\vecx_{m}\left[{m-1\atop 0};R\right]_{1\cdots m-1}}
using the induction hypothesis and Definition~3.1. The last term is also the
$r=0$ term in the desired sum, proving the result for $m$. \endproof

In explicit notation the last proposition reads
\eqn{bincomp}{ (a_{i_1}+x_{i_1})\cdots
(a_{i_m}+x_{i_m})=\sum_{r=0}^{r=m}a_{j_1}\cdots a_{j_r}x_{j_{r+1}}\cdots
x_{j_m}\left[{m\atop r};R\right]^{j_1\cdots j_m}_{i_1\cdots i_m}.}
The main theorem is to actually compute these $R$-binomial coefficient
matrices. We need two lemmas.

\begin{lemma} Denote the {\em monodromy} or `parallel transport' operator in
Definition~3.1 by
\[ \left[a,b;R\right]=(PR)_{a, a+1}\cdots (PR)_{b-1, b}\]
for $a\le b$ with the convention $\left[a,a;R\right]=1$. It obeys
\[ [a,b;R][b,c;R]=[a,c;R]\qquad \forall\ a\le b\le c\]
\[ [a,b;R][c,d;R]=[c+1,d+1;R][a,b;R]\qquad \forall\ a\le c\le d< b\]
\end{lemma}
\proof The proof of the second identity follows from repeated use of the QYBE
or braid relations, in a standard way.
This is easily done by writing the monodromies as braided crossings. Thus
$[a,b;R]$ is the braid of strand $a$ past the strands up to and including
strand $b$, with each crossing represented by an $R$-matrix. If the braid
$[c,d;R]$ lies entirely inside $[a,b;R]$ then it can be pulled through it in
its entirety, giving the commutation relation shown. This is depicted in
Figure~1. \endproof

\begin{figure}
\vskip .3in
\epsfbox{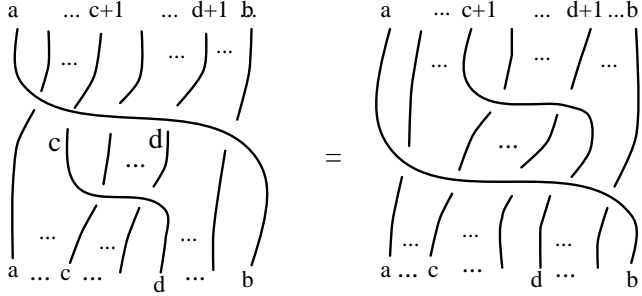}
\caption{Diagram in proof of Lemma~3.3}
\end{figure}

\begin{lemma} The monodromy commutes with the braided-binomial coefficients in
the sense
\[ [1,m;R] \left[{m-1\atop r};R\right]_{1\cdots m-1}= \left[{m-1\atop
r};R\right]_{2\cdots m}[1,m;R]\]
\end{lemma}
\proof We proceed by induction. Thus using Definition~3.1 and the result for
$m-2$ we have
\align{\nqquad [1,m;R] \left[{m-1\atop r};R\right]_{1\cdots m-1}&=&
[1,m;R]\left([r,m-1;R]\left[{m-2\atop r-1};R\right]_{1\cdots m-2}+
\left[{m-2\atop r};R\right]_{1\cdots m-2}\right)\\
&=&[r+1,m;R] [1,m;R]\left[{m-2\atop r-1};R\right]_{1\cdots m-2}+
[1,m;R]\left[{m-2\atop r};R\right]_{1\cdots m-2}\\
&=&[r+1,m;R] \left[{m-2\atop r-1};R\right]_{2\cdots m-1}[1,m;R]+
\left[{m-2\atop r};R\right]_{2\cdots m-1}[1,m;R]}
which equals the right hand side using Definition~3.1 again in reverse. For the
second equality we used the preceding lemma.\endproof

\begin{theorem}
\[ \left[r;R\right]_{1\cdots r}\left[{m\atop r};R\right]_{1\cdots
m}=\left[{m-1\atop r-1};R\right]_{2\cdots m}\left[m;R\right]_{1\cdots m}\]
and hence (formally supposing that all the braided-integers are invertible)
\[ \left[{m\atop r};R\right]_{1\cdots m}=\left[r;R\right]_{1\cdots
r}^{-1}\left[r-1;R\right]^{-1}_{2\cdots r}\cdots \left[2;R\right]^{-1}_{r-1,
r}\left[m-r+1;R\right]_{r\cdots m}\cdots \left[m;R\right]_{1\cdots m}\]
\end{theorem}
\proof We proceed by induction. Suppose the result is true up to $m-1$. Then
from Definition~3.1 and the above lemmas we have
\align{ \left[r;R\right]_{1\cdots r}\left[{m\atop r};R\right]_{1\cdots
m}&&=\left[r;R\right]_{1\cdots r}[r,m;R]
\left[{m-1\atop r-1};R\right]_{1\cdots m-1}+\left[r;R\right]_{1\cdots
r}\left[{m-1\atop r};R\right]_{1\cdots m-1}\\
&&=[r,m;R]\left[r-1;R\right]_{1\cdots r-1}\left[{m-1\atop
r-1};R\right]_{1\cdots m-1}+ [1,m;R]\left[{m-1\atop r-1};R\right]_{1\cdots
m-1}\\
&&\qquad\qquad\qquad+ \left[r;R\right]_{1\cdots r}\left[{m-1\atop
r};R\right]_{1\cdots m-1}\\
&&=[r,m;R]\left[{m-2\atop r-2};R\right]_{2\cdots
m-1}\left[m-1;R\right]_{1\cdots m-1}+\left[{m-1\atop r-1};R\right]_{2\cdots
m}[1,m;R]\\
&&\qquad\qquad\qquad+\left[{m-2\atop r-1};R\right]_{2\cdots
m-1}\left[m-1;R\right]_{1\cdots m-1}\\
&&=\left[{m-1\atop r-1};R\right]_{2\cdots m}[1,m;R]+\left[{m-1\atop
r-1};R\right]_{2\cdots m}\left[m-1;R\right]_{1\cdots m-1}\\
&&=\left[{m-1\atop r-1};R\right]_{2\cdots m}\left[m;R\right]_{1\cdots m}}
as required. Here the first second equality splits $\left[r;R\right]_{1\cdots
r}=\left[r-1;R\right]_{1\cdots r-1}+[1,r;R]$. The $[r,m;R]$ commutes past the
first term of this, while with the second term it combines to give $[1,m;R]$.
The third equality is our induction hypothesis for the outer terms and
Lemma~3.4 for the middle
term. We then used Definition~3.1 in reverse to recognise two of the terms to
obtain the fourth equality. We then recognise $[1,m;R]+[m-1;R]=[m;R]$ to obtain
the result. \endproof

Note that one does not really need the braided-integers to be invertible here
(just as for the usual binomial coefficients). For example, the recursion
relation in the theorem implies that
\eqn{mm-1}{\left[{m\atop m-1};R\right]_{1\cdots
m}=1+(PR)_{m-1,m}+(PR)_{m-1,m}(PR)_{m-2,m-1}+\cdots+(PR)_{m-1,m}\cdots
(PR)_{12}}
which is a right-handed variant of $\left[m;R\right]$. This can also be proven
directly by using the quantum Yang-Baxter equations a lot of times. One can
prove numerous other identities of this type in analogy with usual combinatoric
identities. This theorem demonstrates the beginning of some kind of
braided-number-theory or braided-combinatorics. Because it holds for any
invertible solution of the QYBE, it corresponds to a novel  identity in the
group algebra of the braid group. Physically, it corresponds to `counting' the
`partitions' of a box of braid-statistical particles. These points of view will
be developed elsewhere.

\section{Braided Exponentials and Braided Taylor's Theorem}

As an application of the braided-binomial theorem we are going to study a
braided version of the exponential map.
Throughout this section we assume that $R$ is {\em generic} in the sense that
all the braided-integers $\left[m;R\right]$ are invertible matrices. This
corresponds when defining the usual $q$-exponential to assuming that $q$ is
generic. On the other hand, looking at (\ref{R'}) this means that $R'=P$ for
the corresponding braided vectors and braided covectors. Thus we necessarily
proceed in the case of this free braided differential calculus.
$\Vhaj(P)=k<x_i>$ and $V(P)=k<v^i>$ are free algebras. They have a linear
coproduct making them into braided-Hopf algebras with braiding $\Psi$ defined
by $R$. This seems to be the natural generalization to an $R$-matrix of the
usual 1-dimensional theory of $q$-differential calculus.

\begin{defin} Let $R$ be a generic $R$-matrix. We define the $R$-exponential to
be the formal power-series in $\Vhaj(P)\und\tens\ V(P)$ defined by
\[ \exp_R(\vecx|\vecv)=\sum_{m=0}^{\infty}
\vecx_1\cdots\vecx_m\left[m;R\right]_{1\cdots
m}^{-1}\left[m-1;R\right]_{2\cdots m}^{-1}\cdots \left[2;R\right]_{m-1,
m}^{-1}\vecv_m\cdots\vecv_1.\]
\end{defin}

{}From (\ref{di}) we see at once that
\alignn{efn}{\del^i \exp_R(\vecx|\vecv)& =& \sum_{m=0}^{\infty} \del^i
\vecx_1\cdots\vecx_m\left[m;R\right]_{1\cdots
m}^{-1}\left(\left[m-1;R\right]_{2\cdots m}^{-1}\cdots \left[2;R\right]_{m-1,
m}^{-1}\right)\vecv_m\cdots\vecv_1\nonumber\\
&=& \sum_{m=0}^{\infty} {\bf e}_1^i
\left(\vecx_2\cdots\vecx_m\left[m-1;R\right]_{2\cdots m}^{-1}\cdots
\left[2;R\right]_{m-1, m}^{-1}\vecv_m\cdots\vecv_2\right)\vecv_1\nonumber\\
&=& \exp_R(\vecx|\vecv)v^i}
where $v^i$ is the non-commutative eigenvalue. Recall that the $\del^i$
themselves are a realization of the vector algebra.

\begin{corol} (Braided Taylor's Theorem) in the braided tensor product algebra
$\Vhaj(P)\und\tens \Vhaj(P)$ as in Section~2 we have
\[ \exp_R(\veca|\del) f(\vecx)=f(\veca+\vecx)=\und\Delta f(\vecx).\]
We see that $\del$ is the infinitesimal generator of the translation
corresponding to this braided-coproduct.
\end{corol}
\proof This follows at once from the braided-binomial theorem in Section~3. On
any polymonomial the power-series $\exp_R(\veca|\del)$ truncates to a finite
sum. Computing from (\ref{di}) we have
\align{ && \nqquad \exp_R(\veca|\del)
\vecx_1\cdots\vecx_m=\sum_{r=0}^{r=m}\veca_{1'}\cdots
\veca_{r'}\left[r;R\right]_{1'\cdots r'}^{-1}\cdots
\left[2;R\right]_{r'-1, r'}^{-1}\del_{r'}\cdots\del_{1'}
\vecx_{1}\cdots\vecx_{m} \\
&=&
\sum_{r=0}^{r=m}\veca_{1}\veca_{2'}\cdots\veca_{r'}\left[r;R\right]_{12'\cdots
r'}^{-1}\cdots \left[2;R\right]_{r'-1, r'}^{-1}\del_{r'}\cdots\del_{2'}
\vecx_{2}\cdots\vecx_{m} \\
&=&\cdots\ =\sum_{r=0}^{r=m}\veca_1\cdots\veca_r \left[r;R\right]_{1\cdots
r}^{-1}\cdots \left[2;R\right]^{-1}_{r-1, r} \vecx_{r+1}\cdots\vecx_m
\left[m-r+1;R\right]_{r\cdots m}\cdots \left[m;R\right]_{1\cdots m}\\
&=&(\veca_1+\vecx_1)\cdots (\veca_m+\vecx_m)=\und\Delta(\vecx_1\cdots\vecx_m).}
Here the $1',2'$ etc refer to copies of $M_n$ distinct from the copies labelled
by $1\cdots m$, but they are successively identified by the ${\bf e}^i$ (which
are Kronecker delta-functions) brought down by the application of $\del^i$.
The $ \vecx_{r+1}\cdots\vecx_m$ commute to the right and Theorem~3.5 and
Proposition~3.2 allow us to identify the result. The result is products of
$\und\Delta \vecx$ according to (\ref{covecdelta}) in Section~2. But
$\und\Delta$ is an algebra homomorphism to the braided tensor product, giving
the final form shown. \endproof

\begin{corol} Let $\veca_1\vecx_2=\vecx_2\veca_1R_{12}$ as above. Then
\[:\exp_R(\veca|\vecv)\exp_R(\vecx|\vecv):=\exp_R(\veca+\vecx|\vecv)\]
where $:\ :$ denotes to keep the components of the second copy of $\vecv$ to
the left of the components from the first. The identity takes place as a formal
power-series in $\Vhaj(P)\und\tens \Vhaj(P)\und\tens V(P)$.
\end{corol}
\proof We apply the braided Taylor's theorem just proven,
\align{\exp_R(\veca+\vecx|\vecv)&=& \exp_R(\veca|\del)\exp_R(\vecx|\vecv)\\
&=& \sum_{m=0}^{\infty} \vecx_1\cdots\vecx_m\left[m;R\right]_{1\cdots
m}^{-1}\left[m-1;R\right]_{2\cdots m}^{-1}\cdots \left[2;R\right]_{m-1,
m}^{-1}\exp_R(\vecx|\vecv)\vecv_m\cdots\vecv_1}
using (\ref{efn}). This expression is what we mean by
$:\exp_R(\veca|\vecv)\exp_R(\vecx|\vecv):$. This seems to be the most we can
say in the free case (where there are no commutation relations among the
components of $\vecv$). \endproof

\section{Braided Quantum-Mechanics and Braided Weyl Algebras}

Here we develop a corollary of the fact that the vectors $V(R')$ act on the
covectors $\Vhaj(R')$ by differentiation. Recall that the usual Weyl algebra of
quantum mechanics is the semidirect product of
$x$ by $p$ where $p$ acts on $x$ by differentiation. In our situation with data
$(R,R')$ as in Section~2 we are
in exactly the same situation and hence it is natural to make a (braided)
semidirect product and call the result the braided Weyl algebra for some kind
of braided quantum mechanics. Remarkably, the algebras we recover are just the
ones that have been proposed by other means for this purpose. We understand
here only the algebraic structure of these algebras: suitable $*$-structures
and inner-products are not yet understood in our systematic braided point of
view.

We recall that the data for an ordinary group semidirect product (cross
product) is a group acting on an algebra by automorphisms. One can formulate
the same concept for any Hopf algebra acting on an algebra. Such covariant
systems are called module algebras. There is no problem formulating this just
as easily for a Hopf algebra in a braided category. This is what we need here
and is depicted in Figure~2 for a braided-Hopf algebra $B$ acting on an algebra
$C$. In the diagrammatic notation that we use, the morphisms or maps are
written pointing downwards as nodes joining their inputs to their outputs.
There is a tri-valent vertex for the product in $B$, another for an action
$\alpha:B\tens C\to C$ and another for the coproduct of $B$. We represent
braided transpositions $\Psi$ between any two objects such as $B,C$ by braid
crossings
\eqn{psibra}{\epsfbox{BCpsi.eps}}
There is a symmetry here between positive and negative braid crossings with the
result that we can equally well regard $\Psi^{-1}$ as the braiding of another
braided category with braiding $\bar\Psi$ as shown. Correspondingly, one can
formulate braided module-algebras with either $\Psi$ or $\Psi^{-1}$ in
Figure~2. If our braided-coproduct $B\to B\und\tens B$ is an
algebra-homomorphism to the braided tensor product with $\Psi$ as in
(\ref{bratens}) it is appropriate to use the positive braid with $\Psi$ in
Figure~2. We have shown it for the negative braid because this is the one which
our example will satisfy.

\begin{figure}\vskip .3in
\qquad\qquad\epsfbox{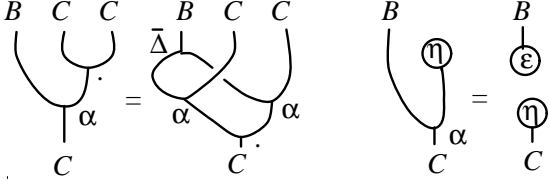}
\caption{The notion of braided-module algebra (with reversed braiding). The
braided-covectors are such a braided-module algebra under the braided-vectors
(with reversed braiding).}
\end{figure}

Firstly, we denote the generators of $V(R')$ by $p^i$ rather than $v^i$ in
Section~2 (because they are going to become momentum) and we define $B=\bar
V(R')$ to be the same as this $V(R')$ and the same coproduct on the generators
but reversed braiding
\eqn{deltacop}{\bar\Delta(p^i)=p^i\tens 1+1\tens
p^i,\quad\bar\Psi(\vecp_1\tens\vecp_2)=R_{21}^{-1}\vecp_2\tens\vecp_1}
when it comes to extending to products. By definition $\bar\Delta$ extends as
an algebra homomorphism $B\to B\bar{\tens} B$ where the latter is defined as in
(\ref{bratens}) but with $\bar\Psi=\Psi^{-1}$ in the role of $\Psi$. It is an
equally good braided-Hopf algebra but lives in the category with reversed
braiding from our original $V(R')$. From now on we do everything in this
category.

\begin{lemma} $\Vhaj(R')$ is a braided $\bar{V}(R')$-module algebra in the
sense of Figure~2, with action $\alpha$ defined by the operators $p^i\la
a=\del^i(a)$.
\end{lemma}
\proof This follows in fact from general principles as outlined in
\cite{Ma:introp} to do with the identification of
$\bar{\Delta}=\Psi^{-1}\circ\und\Delta$. It can also be seen by iterating
Lemma~2.3. Firstly $\bar{\Delta}\vecp_1\cdots\vecp_r$ is given by braided
binomial coefficient-matrices similar to those we have seen in Section~3 for
braided covectors. For example,
\eqn{deltacoppp}{ \bar{\Delta}\vecp_1\vecp_2=\vecp_1\vecp_2\tens
1+1\tens\vecp_1\vecp_2+\vecp_1\tens\vecp_2+R_{21}^{-1}\vecp_2\tens\vecp_1.}
Apply this in the right hand side in Figure~2 gives in this case four terms. On
the other hand computing the action of $\vecp_1\vecp_2$ on the left hand side
directly by $\del_1\del_2$ and knowing that these $\del$ are each derivations
from Lemma~2.3 also gives four terms. Comparing them and using  functoriality
of the braiding one can see that they are equal. Similarly the general form
follows from the Leibniz rule for each $\del$. \endproof

\begin{example} In the 1-dimensional case of Example~2.5 the module-algebra
property corresponds to the identity
\[ \sum_{s=0}^{r} \left[{r\atop s};q^{-1}\right]\left[{a-r\atop
b-s};q\right]q^{b(r-s)}=\left[{a\atop b};q\right],\quad \forall\ r,b\le a.\]
for $q$-binomial coefficients. These are considered zero outside the usual
range.
\end{example}
\proof We examine the braided module-algebra property directly in this
1-dimensional setting. For the left in Figure~2 we have
\eqn{pmn}{p^r\la(x^{n+m})=[n+m;q]\cdots[n+m-r+1;q]x^{n+m-r}}
where $\la$ denotes the action $\alpha$. On the right hand side by contrast we
have, using the usual  $q$-binomial theorem to compute $\bar{\Delta}p^r$, the
expression
\align{&&\nquad \sum_{s=0}^{r} \left[{r\atop
s};q^{-1}\right]p^r\la\bar\Psi(p^{r-s}\tens x^n)\la x^m\\
&&=\sum_{s=0}^{r} \left[{r\atop s};q^{-1}\right] (p^s\la x^n)(p^{r-s}\la
x^m)q^{n(r-s)}\\
&&=\sum_{s=0}^{r} \left[{r\atop s};q^{-1}\right][n;q]\cdots
[n-s+1;q][m;q]\cdots[m-r+s+1;q]q^{n(r-s)}x^{n+m-r}}
We see that the module algebra condition for all $a=n+m$ and $b=n$ corresponds
to the identity shown. The $q^{n(r-s)}$ here comes from the braiding of $p$
past $x$. \endproof

Now whenever we are in the situation of Figure~2 with $C$ a braided $B$-module
algebra, we can make a semidirect product\cite{Ma:bos}. This is built on the
object $C\tens B$ but with the multiplication twisted by the action $\alpha$ as
shown in Figure~3.

\begin{figure}\vskip .3in
\qquad\qquad\epsfbox{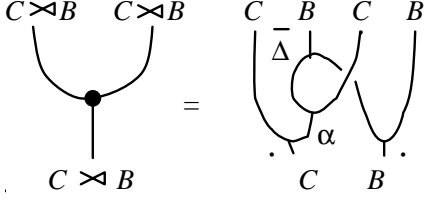}
\caption{Braided Cross-Product $C\cocross B$. The braided Weyl algebra has this
form.}
\end{figure}

\begin{propos} The {\em braided Weyl algebra} $\Vhaj(R')\cocross \bar{V}(R')$
defined as the braided semidirect product by  the action of braided vectors on
braided covectors by $\del^i$ is generated by $\hat\vecx=\vecx\tens 1$ in $
\Vhaj(R')$ and $\hat{\vecp}=1\tens\vecp$ in $V(R')$ with
cross relations
\[ \hat{\vecp_1}\hat{\vecx_2}-\hat{\vecx_2}R_{21}\hat{\vecp_1}=\id.\]
\end{propos}
\proof We can compute the expression for the product of a general element
$\vecx_1\cdots\vecx_r\tens\vecp_{r+1}\cdots\vecp_m$ with another such element
from Figure~2. One uses the braided-binomial theorem to compute
$\bar{\Delta}\vecp_{r+1}\cdots\vecp_m$. To know only the relations of
the resulting algebra we can concentrate on $\hat{\vecp}$ and $\hat{\vecx}$ as
stated. On the one hand we have
$(\vecx_2\tens1)(1\tens\vecp_1)=\vecx_2\tens\vecp_1$ because $\bar{\Psi}$ is
trivial on $1$. On the other hand we have
\eqn{crossRheis}{(1\tens\vecp_1)(\vecx_2\tens1)=
\vecp_1\la\vecx_2+\bar{\Psi}(\vecp_1\tens\vecx_2)=
\id+\vecx_2\tens R_{21}\vecp_1}
giving the commutation relations shown. The braiding is as in Lemma~2.3 and
taken is from \cite{Ma:lin}. Note that we do not begin with such relations and
afterwards verify that
they are consistent, but rather we have a well-defined algebra on the tensor
product space from the start. \endproof

Some similar commutation relations in the $SL_q(n)$ case were studied in
\cite{Kem:sym} as a kind of $q$-quantization  of harmonic oscillators, but our
construction works more generally for any $R$ with associated $R'$ ($R$ need
not be of Hecke type). They are also familiar in the theory of quantum inverse
scattering in various contexts. The main point however, is that the
construction is understood now in a systematic way as a braided-semidirect
product.

\begin{example} Let $R=(q)$ the 1-dimensional $R$-matrix as in Example~2.5. The
braided semidirect product $k[x]\cocross k[p]$ where $p\la f(x)=\del f(x)$ (the
$q$-derivative) is the algebra generated by $\hat p=1\tens p$ and $\hat
x=x\tens 1$ and relations
\[ \hat p\hat x-q\hat x\hat p=1.\]
\end{example}
\proof The proof for the commutation relations is just as the previous example.
One can also compute the product of  general basis elements in closed form
using the $q$-binomial theorem, thus
\alignn{pf}{\hat{p}^mf(\hat{x})&=&(1\tens p^m)(f(x)\tens 1)=\sum_{r=0}^m
\left[{m\atop r};q^{-1}\right] p^r\la \Psi(p^{m-r}\tens f(x))\nonumber\\
&&\nqquad =\sum_{r=0}^m \left[{m\atop r};q^{-1}\right](\del)^r f(q^{m-r}x)\tens
p^{m-r}=\sum_{r=0}^m \left[{m\atop r};q^{-1}\right](\del^r
f)(q^{m-r}\hat{x})q^{r(m-r)}\hat{p}^{m-r}.}
For example, $\hat{p}f(\hat{x})=f(q\hat{x})\hat{p}+(\del f)(\hat{x})$.
\endproof

This is the $q$-Heisenberg algebra as proposed for $q$-deformed quantum
mechanics in \cite{Man:not}\cite{SchWes:def}. We see that it too is a
semidirect product exactly of the type of the usual Weyl algebra. That it is
built as we have stated on $k[x]\cocross k[p]$ followed from the non-trivial
fact that $\del$ extended to products as a module-algebra action. The main
point of our approach is that all formulae follow as in the usual un-deformed
case provided only that we remember the braid-statistics introduced by $q$ or
$R$. For example, the $q$-Schroedinger representation of this algebra can be
written down immediately along the usual lines, as in principle can other
desiredata of (deformed) quantum mechanics. This will be explored elsewhere.

\end{document}